\documentclass[draft,showkeys,noshowpacs,eqsecnum,nofootinbib,aps]{revtex4}

\def\beq{\begin{equation}}
\def\eeq{\end{equation}}
\def\bea{\begin{eqnarray}}
\def\eea{\end{eqnarray}}\def\nn{\nonumber}

\def\na{\nabla}
\def\pa{\partial}

\def\nn{\nonumber}

\begin{document}
\title{Equation of State Parameters for Fluid of Stringy Extended Objects in Cosmology with Cosmological Constant}
\author{Soon-Tae Hong}
\email{galaxy.mass@gmail.com} 
\affiliation{Department of Physics, Jeonbuk National University, Jeonju 54896, South Korea}
\author{Bum-Hoon Lee}
\email{bhl@sogang.ac.kr} 
\affiliation{Center for Quantum Spacetime, Sogang University, Seoul 121-742 South Korea}
\affiliation{Department of Physics, Sogang University, Seoul 121-742 South Korea}
\affiliation{Department of Physics, Shanghai University, Shanghai 200444 China}
\date{January 22, 2003}
\date{\today}
\begin{abstract}
We construct the strong energy conditions (SECs) for both massive and massless stringy extended objects in the higher dimensional cosmology (HDC) with 
cosmological constant $\Lambda$. Exploiting these conditions, we find the equation of state (EoS) parameters \mbox{$w\geq -(D-4)/D$} 
for both the massive and massless stringy extended objects in $D$ {$(D\geq 5)$} dimensional cosmology. The stringy SECs impose a universal constraint on $w$ that remains valid across both radiation- and matter-dominated eras. We elucidate the relations between the EoS parameter in 
the HDC with cosmological constant and that of Hawking--Penrose limit for the massive and massless point particles in the 
four dimensions. 
We evaluate the EoS parameters in terms of the contributions from the point particle property, 
cosmological constant, and extended object degrees of freedom, respectively. We also investigate the weak energy condition for the massive and massless stringy extended objects in the HDC, and those for the massive and massless point particles in the four dimensions,~respectively. 
\end{abstract}
\maketitle

\section{Introduction}

The observational data for the accelerating expansion of the universe has suggested a positive vacuum expectation value of the cosmological 
constant~\cite{per98}. In the standard Big Bang model, it is believed that, after the Big Bang
explosion, the radiation-dominated era occurred, followed by the matter-dominated one, even though there was a
hot thermalization period of radiation and matter immediately after the Big Bang. In the standard cosmology, the
equation of state (EoS) of the fluid of massive point particles is different from that of the fluid of massless point particles. 

The standard Big Bang cosmology is based on the Hawking--Penrose singularity theorem, which relies 
on the strong energy condition (SEC)~\cite{hawking70}. 
On the other hand, in the four-dimensional point particle cosmology~\cite{hawking73,wald84,hobson06,hong08,hong11}, a perfect fluid is introduced to 
describe a continuous distribution of matter with energy-momentum tensor $T_{ab}$ $(a, b =0,1,2,3)$  
in terms of the mass-energy density and the pressure. 
This fluid is called perfect because of the absence of heat conduction terms and stress terms corresponding to viscosity~\cite{wald84}. 
The $D$ dimensional perfect fluid property has been applied to 
the higher dimensional cosmology (HDC) without cosmological constant~\cite{hong08,hong11}.

In this paper, 
we call the fluid in $D=4+p$ $(D\geq 5)$ dimensional manifold consisting of extended objects of the $p$-branes, the fluid of stringy extended objects, for simplicity. To be more specific, in $D=5$ dimensions, the stringy extended objects denote the $p=1$ stringy objects (or stringy particles) having one-dimensional fiber~\cite{hong08,hong11}. Note that while the theoretical bounds on the EoS parameter 
$w_{\rm total}$ 
 for the 
fluid of radiation (fluid of massless stringy extended  objects) and fluid of matter (fluid of massive stringy extended  extended objects) 
are continuous, the physical dominance of species still evolves. 
Note also that in the point particle cosmology, there are 
two EoSs for the fluids massless and massive particle eras, respectively. In the HDC, 
as the early universe evolves with the expansion rate, this rate increases and the 
twist of the fluid of stringy extended objects decreases exponentially. The effect of the shear has been shown to be negligible and thus the universe is considered 
 isotropic and homogeneous~\cite{hong08,hong11}. 
 
 Since Hubble discovered the expansion of our universe, the Big Bang cosmology has been developed into a precision science by 
cosmological observations including supernova data~\cite{per98} and measurements 
of cosmic microwave background (CMB) radiation~\cite{spergel07}. These observations triggered an explosion of recent 
interest in the origin of dark energy~\cite{per992}.

The local measurements of Hubble constant $H_{0}$ are in tension with early universe measurements such as CMB and 
baryon acoustic oscillation (BAO), within the standard cosmological constant cold dark matter ($\Lambda$CDM) 
 model. 
Recently, the holographic Friedman--Lema\^{i}tre--Robertson--Walker (FLRW) universe and the ensuing observational 
constraints were investigated on the four-dimensional membrane embedded in the $D=4+1$ dimensional bulk 
spacetime~\cite{bhl21,cai18,cai19}. In the HDC without background of cosmological constant, the singularities in geodesic surface congruence for the time-like 
and null stringy extended  particles were investigated to yield the Raychaudhuri-type equations possessing correction terms related with the 
characteristics owing to the stringy extended  objects~\cite{hong08,hong11}. Assuming the stringy SEC in the HDC without the cosmological constant, 
the Hawking--Penrose-type EoS inequality equations were obtained. To be specific, 
the stringy SECs of both the fluids of massive and massless stringy extended objects produce the same EoS inequality in the HDC without the 
cosmological constant. This EoS inequality has been shown to be not equivalent to the Hawking--Penrose EoS inequality equations in the 
four-dimensional FLRW cosmology~\cite{hong08,hong11}. Recently, 
 the null energy condition \cite{carroll} 
(here, the weak energy condition (WEC) for the massless point particles
) has been applied to the models of dynamical black holes~\cite{ion25,erik24}. 

In this paper, we 
 investigate the SECs, WECs and the ensuing EoS in the HDC in the background of cosmological constant by 
extending the results of the studies
~\cite{hong08,hong11}. To 
this end, we 
study singularities associated with the Raychaudhuri-type equations and construct the SECs and the ensuing EoSs in geodesic surface congruence 
in the HDC. We 
then discuss the cosmological constant in the HDC. Moreover, we 
elucidate the origin of the relations between the 
Hawking--Penrose-type EoS inequality equations in the four-dimensional limits in the HDC, which are missing in Refs.~\cite{hong08,hong11}, 
and the Hawking--Penrose EoS inequality in the four dimensions in the FLRW~cosmology. 

Next, we 
explicitly construct the SECs for the fluid of stringy extended  objects in the HDC having the cosmological constant contributions, to show 
that the EoS inequalities for the fluid of massive stringy extended  objects are the same as those for the massless ones, even though their corresponding formulas 
are different to each other. Moreover, in our study, we  
evaluate the energy conditions for the fluid of point particles 
possessing the mass-energy density $\rho_{0}$ and the pressure $P_{0}$ described in the four-dimensional cosmology. 

Note that in this paper, we  
introduce a fibration, $\pi:M\rightarrow N$ over a base manifold $N$ (spanned in total manifold $M$)
associated with a fiber space $F$. Here,   
the base manifold $N$ is fixed to reside on the four-dimensional spacetime and the stringy effects are considered in the total $D$ ($D\geq 5$) 
dimensional manifold of the fiber bundle. {To be more specific, the HDC is defined in the $D$ ($D\geq 5$) dimensional fiber bundle manifold. 
By definition, the HDC is thus different from the four-dimensional cosmology which does not possess $F$.} 
Moreover, in the HDC, the radiation in radiation-dominated era is described in terms of the fluid of massless stringy extended objects, for 
instance~\cite{hong08,hong11}. Similarly, in the HDC, the matter in a matter-dominated era is delineated in terms of the fluid of  massive stringy 
extended  objects. The fluid of massive or massless point particles is then defined in the four-dimensional spacetime which does not possess 
$F$~\cite{hong08,hong11,steenrod51}. Next, 
 the cosmological constant $\Lambda$ is introduced in the background.

In Section 2, we 
 construct the geometry and Raychaudhuri-type equations in 
the HDC. We 
sketch the HDC with cosmological constant. In Section 3, we 
 formulate the SECs, WECs and EoS parameters for the fluid of massive stringy extended objects and discuss the cosmological constant in the HDC. 
In Section 4, we 
 formulate the SECs, WECs and EoS parameters for the fluid of massless stringy extended 
objects and study the cosmological constant in the HDC. 
Section 5 includes conclusions. In Appendix \ref{app1}, we 
 pedagogically study four-dimensional point particle cosmology. In Appendix \ref{app2}, we will study the details of the Raychaudhuri-type equation in the HDC. 
In particular, we 
list the definitions of the variables employed in Section 2.

\section{Geometry and Raychaudhuri-Type Equations in HDC}\label{section2}
\renewcommand{\theequation}{\arabic{equation}}

In this Section, we investigate the geometry of the HDC and the ensuing Raychaudhuri-type equations. To this end, we first introduce a fibration, 
$\pi:M\rightarrow N$ over a base manifold $N$ associated with $F$~\cite{steenrod51,frankel97,hong11}. 
In this paper, 
 the base manifold $N$ is fixed to reside on the four-dimensional spacetime.
In analogy of the relativistic action of point particles in $N$, the action for stringy extended object is proportional to the area of the surface 
spanned in total manifold $M$ by the evolution along time direction of the stringy extended object in $F$.
Note that in the HDC defined in the $D=4+p$ $(p\geq 1)$ manifold, the extended manifold is associated with the extended $p$-brane.
The $(p=1)$-string is the stringy extended object which resides on the five-dimensional manifold possessing the one-dimensional $F$. Note also that the point particle 
lives on the four-dimensional manifold without $F$.
In order to define the action on the curved manifold, let $(M, g_{ab})$ be a
$D$ dimensional manifold associated with the metric $g_{ab}$.
Given $g_{ab}$,  there can be 
 a unique covariant derivative $\na_{a}$ satisfying~\cite{wald84} $\na_{a}g_{bc}=0$,
$\na_{a}\omega^{b}=\pa_{a}\omega^{b}+\Gamma^{b}_{~ac}~\omega^{c}$ (with $\pa_{a} \equiv \pa/\pa{x_a}$ and $\Gamma^{b}_{~ac}$ being the Christoffel symbol) 
and $(\na_{a}\na_{b}-\na_{b}\na_{a})\omega_{c}=R_{abc}^{~~~d}~\omega_{d}$ $(a=0,1,...,D-1)$, with the curvature $R_{abc}^{~~~d}$. 
Next we investigate the Raychaudhuri-type equations which are 
studied in terms of the SECs for the fluid of massive and 
massless stringy extended objects and the point particles with cosmological constant. 
Moreover the Raychaudhuri-type equations will be discussed to clarify the attractive gravitional force.

We investigate explicitly the action and the ensuing geometry in the HDC. To 
this end, 
Appendix \ref{app1} ~first studies 
the total action, which consists of the relativistic action for the point particle and actions of gravity and perfect fluid, for 
the four-dimensional point particle cosmology having the cosmological constant $\Lambda$ in the curved manifold. 
Keeping the total action for the point particles in mind, we investigate the stringy extended  objects in the curved 
manifold. To this end, one needs 
 to introduce the gravity action in addition to the extended $(p=1)$-brane action 
(or Nambu--
Goto action)~\cite{nambu70,goto71}. In this paper, 
 we investigate 
the HDC possessing the cosmological constant $\Lambda$ and the ensuing EoS. The case without cosmological constant has been analyzed in 
Refs.~\cite{hong07,hong08,hong11}. We start with $D$ dimensional gravity related with total action of the form 
\bea
S&=&S_{p=1}+S_{\rm gr}+S_{\rm pf},\nn\\
S_{p=1}&=&-\kappa\int_{\tau_{1}}^{\tau_{2}}d\tau \int_{0}^{\pi}d\sigma f(\tau,\sigma),\nn\\
S_{\rm gr}&=&\frac{1}{16\pi}\int d^{D}x\sqrt{-g}(R-2\Lambda),
\label{sss2}
\eea
where $S_{p=1}$, $S_{\rm gr}$ and $S_{\rm pf}$ are the extended $(p=1)$-brane action, gravity and perfect fluid actions, respectively,  
$\kappa=\frac{1}
{2\pi\alpha^{\prime}}$ with $\alpha^{\prime}$ being 
the universal slope of Regge trajectory and $\Lambda$ is a cosmological constant, $\tau$ and $\sigma$ are the world sheet coordinates, $R$ is 
the Ricci scalar, and $g$ is the determinant of the metric tensor. 
Note that  $D$ dimensional cosmological constant is given by 
$\Lambda=[(D-1)(D-2)/6]\Lambda_{0}$ where $\Lambda_{0}$ is the (3 + 1)-dimensional cosmological constant associated with the point 
particle~\cite{hong08ph}. 
In the fiber bundle formalism~\cite{hong08,hong11,steenrod51}, the extra dimension is defined to stand for the dimension 
of the fiber space where the (massive or massless) stringy extended objects having size reside. 
Here, the fiber space describes the geometry of a single stringy extended  particle. Moreover, $\Lambda_{0}$ which is related to the observed 
vacuum energy is given by $\Lambda_{0}\simeq 1.1\times 10^{-52}$ m$^{-2}$.

Next we investigate the Nambu--Goto action $S_{p=1}$  in Equation 
(\ref{sss2}) associated with $f(\tau,\sigma)$~\mbox{\cite{scherk75,hong07}}. 
To 
 this end, we digress to note that the point particles have coordinates in the four-dimensional base manifold $N$, 
which is described in the action $S_{p=0}$ in Equation (\ref{sss})  having $f(\tau)=\left(-g_{ab}\frac{\pa x^{a}}{\pa\tau}
\frac{\pa x^{b}}{\pa\tau}\right)^{1/2}=(-g_{ab}\xi^{a}\xi^{b})^{1/2}$. 
Here, we have $\xi^{a}=(\pa/\pa\tau)^{a}$ 
$(a=0,1,2,3)$ with metric $g_{ab}$ associated with $(-,+,+,+)$~\cite{scherk75,hong07,gova}. Moreover, the variation of $S_{\rm gr}+S_{\rm pf}$ under the 
metric change $\delta g^{ab}$ $(a,b=0,1,2,3)$ 
affects the Einstein field equation (\ref{einfieldeq2})
.

{In contrast, the stringy extended objects have the coordinates in the $D$ $(D\geq 5)$ dimensional total manifold $M$, 
which is delineated in the action $S_{p=1}$ in Equation (\ref{sss2}) possessing $f(\tau,\sigma)$ defined as 
\beq
f(\tau,\sigma)=(-\det h_{MN})^{1/2}=[(\xi\cdot\zeta)^{2}-(\xi\cdot\xi)(\zeta\cdot\zeta)]^{1/2},
\label{ftausigma}
\eeq
where 
the relation $h_{MN}=g_{ab}\frac{\pa x^{a}}{\pa\sigma^{M}}\frac{\pa x^{b}}{\pa\sigma^{N}}$, with $\sigma^{M}=(\tau,\sigma)$ $(M=0,1)$,
and the notations $\xi^{a}=(\pa/\pa\tau)^{a}$ and $\zeta^{a}=(\pa/\pa\sigma)^{a}$ $(a=0,1,...,D-1)$ are used. 
Here, the metric $g_{ab}$ $(a=0,1,...,D-1)$ is related with $(-,+,\cdots,+)$. 
Note that in this paper, 
for completeness, in addition to $S_{p=1}$ we included the actions $S_{\rm gr}$ and $S_{\rm pf}$, which were not treated in the 
Refs.~\cite{hong07,hong08,hong11}. To be more specific, the variation of $S_{\rm gr}+S_{\rm pf}$ under $\delta g^{ab}$ \mbox{$(a,b=0,1,...,D-1)$} 
affects the Einstein field 
equation of the form: \mbox{$R_{ab}-\frac{1}{2}g_{ab}R+g_{ab}\Lambda=8\pi T_{ab}$} \mbox{$(a,b=0,1,...,D-1)$} for a perfect fluid. 
Here, $R_{ab}$ and $R$ are 
the Ricci curvature tensor and the Ricci scalar curvature, respectively.}

We consider a smooth congruence of time-like geodesic surfaces in
$M$. We parameterize the surface generated by the evolution of a
time-like string by 
~$\tau$ and $\sigma$, and then there are 
the corresponding vector fields
$\xi^{a}=(\pa/\pa\tau)^{a}$ and $\zeta^{a}=(\pa/\pa\sigma)^{a}$ as shown in Equation 
(\ref{ftausigma})~\cite{hong11,scherk75,hong07}. Note that $\xi^{a}$ is a time-like vector field and 
$\zeta^{a}$ is a space-like one. Since we have gauge degrees of freedom (DOF), we can choose the
orthonormal gauge~\cite{scherk75,hong07,hong11} 
\beq
\xi\cdot\zeta=0,~~~\xi\cdot\xi+\zeta\cdot\zeta=0.
\label{orthogauge}
\eeq
where the plus sign in the second equation is due to the 
feature
that $\xi\cdot\xi$ is timelike and $\zeta\cdot\zeta$ is spacelike. Note that the gauge fixing 
(\ref{orthogauge}) for the world sheet coordinates 
signifies 
that the tangent vectors are orthonormal everywhere up to a local scale factor~\cite{scherk75,hong07}.

We introduce the deviation vector field $\eta^{a}=(\pa/\pa
\alpha)^{a}$ which represents the displacement to an
infinitesimally nearby world sheet, and let $\Sigma$ denote the
three-dimensional submanifold spanned by the world sheets  
$\gamma_{\alpha}(\tau,\sigma)$. One 
then may choose $\tau$,
$\sigma$ and $\alpha$ as 
coordinates of $\Sigma$ to yield the
commutator relations~\cite{hong07,hong11}
\beq
\pounds_{\xi}\eta^{a}=\xi^{b}\na_{b}\eta^{a}-\eta^{b}\na_{b}\xi^{a}=0\ \ 
\text{and}\ \ 
 \pounds_{\zeta}\eta^{a}=\zeta^{b}\na_{b}\eta^{a}-\eta^{b}\na_{b}\zeta^{a}=0.
\label{poundxizeta}
\eeq

Exploiting 
 the 
 commutators (\ref{poundxizeta}), 
one arrives 
 at
\beq
\frac{dS_{p=1}}{d\alpha}=-\int_{\tau_{1}}^{\tau_{2}} d\tau \int_{0}^{\pi}d\sigma (P_{\tau}^{b}\xi^{a}\na_{a}\eta_{b}
+P_{\sigma}^{b}\zeta^{a}\na_{a} \eta_{b})
=\int_{\tau_{1}}^{\tau_{2}}d\tau \int_{0}^{\pi}d\sigma\eta_{b}(\xi^{a}\na_{a}P_{\tau}^{b}+\zeta^{a}\na_{a}P_{\sigma}^{b})+{\rm boundary~terms},\label{dsng0}
\eeq
where we introduced  the energy-momentum currents
\beq
P_{\tau}^{a}=\frac{\kappa}{f}[(\xi\cdot\zeta)\zeta^{a}-(\zeta\cdot\zeta)\xi^{a}]
\ \ \text{and}\ \ P_{\sigma}^{a}=\frac{\kappa}{f}[(\xi\cdot\zeta)\xi^{a}-(\xi\cdot\xi)\zeta^{a}].
\label{currentss}
\eeq
Here, the boundary terms vanish if one exploits 
the boundary conditions: $\eta^{a}(\tau=\tau_{1}; \sigma)=\eta^{a}(\tau=\tau_{2}; \sigma)=0$ 
and $P_{\sigma}^{a}(\tau; \sigma=0)=P_{\sigma}^{a}(\tau; \sigma=\pi)=0$~\cite{scherk75,gova} to produce a stringy geodesic surface equation: 
$\xi^{a}\na_{a}P_{\tau}^{b}+\zeta^{a}\na_{a} P_{\sigma}^{b}=0$. Applying the orthonormal gauge conditions to the above stringy geodesic surface equation, 
one obtains 
the stringy geodesic surface equation of the form
\beq
-\xi^{a}\na_{a}\xi^{b}+\zeta^{a}\na_{a}\zeta^{b}=0.
\label{gEOSe}
\eeq

Exploiting the first equation in Equation 
(\ref{dsng0}), one is 
left with the second derivative of $S_{p=1}$ to yield a stringy geodesic deviation 
equation~\cite{hong07}
(for 
 more algebraic details associated with outline derivations of 
${dS_{p=1}}/{d\alpha}$ and ${d^{2}S_{p=1}}/{d\alpha^{2}}$, 
see Appendix \ref{app1}, where these two quantities are pedagogically explained in the point particle case)
\beq
\frac{d^{2}S_{p=1}}{d\alpha^{2}}=\int_{\tau_{1}}^{\tau_{2}} d\tau\int_{0}^{\pi} d\sigma\eta_{a}(\chi\eta)^{a},
\label{gdev1}
\eeq
where, naively 
 speaking, the second derivative 
${d^{2}S_{p=1}}/{d\alpha^{2}}$  
is related with the curvature 
$R_{bcd}^{~~~a}$ in the $D$ $(D\geq 5)$ dimensional HDC: 
\beq
(\chi\eta)^{a}=\xi^{b}\na_{b}(\eta^{c}\na_{c}P_{\tau}^{a})+\zeta^{b}\na_{b}(\eta^{c}\na_{c}P_{\sigma}^{a})+R_{bcd}^{~~~a}
(\xi^{b}P_{\tau}^{d}+\zeta^{b}P_{\sigma}^{d})\eta^{c}, 
\label{aleph2}
\eeq
which is then applied to the $D$ dimensional Einstein field equation 
. 

On the other hand, the variation in $S_{\rm gr}+S_{\rm pf}$ under the change $\delta g^{ab}$ ($a,b=0,1,...,D-1$) yields the Einstein field equation
\beq
R_{ab}-\frac{1}{2}g_{ab}R+g_{ab}\Lambda=8\pi T_{ab},
\label{eineq}
\eeq
where 
the definition $R_{ab}=R_{acb}^{~~~c}$ is exploited 
with $R_{bcd}^{~~~a}$ $(a=0,1,...,D-1)$ being defined in Equation (\ref{aleph2}).
Here, we exploited the relation 
\beq
\frac{\delta S_{\rm pf}}{\delta g^{ab}}=-\frac{1}{2}\sqrt{-g}T_{ab}.
\label{spfdel}
\eeq 
Note that the stringy effect in $S_{p=1}$ in Equation (\ref{sss2}) affects 
the $D$ ($D\geq 5$) dimensional Einstein field equation 
 (\ref{eineq}). 
Note also that $R_{ab}$ in Equation (\ref{eineq}) is affected by the cosmological constant $\Lambda$. 
In this study, 
the cosmological constant $\Lambda$ is introduced to 
investigate the EoS associated with the non-zero and positive $\Lambda$ in Section 3.

We briefly investigate the HDC associated with the Raychaudhuri-type equation for the fluid of massive stringy extended 
objects~\cite{hong08,hong11}: 
\beq
\frac{d\theta}{d\tau}-\frac{d\bar{\theta}}{d\sigma}=-\frac{1}{D-1}(\theta^{2}-\bar{\theta}^{2})-\sigma_{ab}\sigma^{ab}+\bar{\sigma}_{ab}\bar{\sigma}^{ab}
+\omega_{ab}\omega^{ab}-\bar{\omega}_{ab}\bar{\omega}^{ab}-R_{ab}(\xi^{a}\xi^{b}-\zeta^{a}\zeta^{b}).
\label{rayeqn}
\eeq
Here, $(\theta,\sigma_{ab},\omega_{ab})$ are the expansion, shear and twist of the universe and $(\bar{\theta},\bar{\sigma}_{ab},\bar{\omega}_{ab})$ 
are those of the string, respectively. The more details of these variables are 
defined in Appendix \ref{app2}. To be specific, in the point particle case, 
there are 
only DOF of $(\theta,\sigma_{ab},\omega_{ab})$ and $\xi^{a}$. In contrast, in the HDC we possess new DOF 
related with the physical variables $(\bar{\theta},\bar{\sigma}_{ab},\bar{\omega}_{ab})$ and $\zeta^{a}$. In particular, in the HDC, 
there are 
the DOF of the twist (or rotation) of the string associated with $\bar{\omega}_{ab}$. 
This feature implies that the string performs the dynamic rotational motion with respect to the four-dimensional base manifold $N$. 
For more derivation details of the Raychaudhuri-type equation 
 (\ref{rayeqn}), see Appendix \ref{app2}. 
We assume that $\sigma_{ab}=\bar{\sigma}_{ab}$ and $\omega_{ab}=\bar{\omega}_{ab}$. Taking an ansatz that the expansion $\bar{\theta}$ is constant 
along the $\sigma$ direction, one ends 
 up with 
\beq
\frac{d\theta}{d\tau}=-\frac{1}{D-1}(\theta^{2}-\bar{\theta}^{2})-R_{ab}(\xi^{a}\xi^{b}-\zeta^{a}\zeta^{b}).
\label{raych}
\eeq

We assume an SEC for the fluid of massive stringy extended  particles:  
\beq
R_{ab}(\xi^{a}\xi^{b}-\zeta^{a}\zeta^{b})\ge 0,
\label{sec}
\eeq
then the Raychaudhuri-type equation in (\ref{raych}) has a solution of the form
\beq
\frac{1}{\theta(\tau)}\geq \frac{1}{\theta(0)}+\frac{1}{D-1}\left(\tau-\int_{0}^{\tau}d\tau\left(\frac{\bar{\theta}}{\theta}\right)^{2}\right).
\label{soln1}
\eeq

We assume that $\theta(0)$ is negative so that the congruence is initially converging as in the point particle case~\cite{hawking70}. The EoS 
inequality 
(\ref{soln1}) then implies that $\theta(\tau)$ should pass through the singularity within a proper time
\beq
\tau\leq\frac{D-1}{|\theta(0)|}+\int_{0}^{\tau}d\tau\left(\frac{\bar{\theta}}{\theta}\right)^{2}.
\eeq

Similar to the massive stringy extended  particle case in Equation 
(\ref{raych}), we consider the Raychaudhuri-type equation associated with expansion of the fluid of 
massless stringy extended objects related with the corresponding null vector $k^{a}$~\cite{hong08,hong11}: 
\beq
\frac{d\theta}{d\lambda}=-\frac{1}{D-2}\theta^{2}+\frac{1}{D-1}\bar{\theta}^{2}-R_{ab}(k^{a}k^{b}-\zeta^{a}\zeta^{b}),
\label{raych2}
\eeq
where $\lambda$ is an affine parameter. {In this paper, 
we consider 
an ansatz that the photon corresponding 
to the stringy extended particle associated with the null vector $k^{a}$ and the affine parameter $\lambda$ is approximately massless, for 
simplicity~\cite{hong08,hong11}. ({{In 
 the conventional string 
theory, there are 
massless photon in the open string sector and massless graviton in the closed string one
~\cite{scherk75,gova}}}.)
Note that there is 
the upper limit of 
exceptionally 
small experimental value} for the photon mass 
$M(\text{photon})=1.00 \times 10^{-27}$ GeV~\cite{pdg}. 
For more details on $k^{a}$ and $\lambda$,~see Section 4. 
Note also that, differently from the factor $D-1$ in Equation (\ref{raych}), there is 
the factor $D-2$ in the first term in Equation (\ref{raych2}) which originates from the property 
that the massless particle has only two transverse DOF without a 
longitudinal DOF. We assume an SEC for the fluid of massless stringy extended  particles: 
\beq
R_{ab}(k^{a}k^{b}-\zeta^{a}\zeta^{b})\ge 0,
\label{sec2}
\eeq
then the Raychaudhuri-type equation 
(\ref{raych2}) has a solution
\beq
\frac{1}{\theta(\lambda)}\geq \frac{1}{\theta(0)}+\frac{1}{D-2}\left(\lambda-\frac{D-2}{D-1}
\int_{0}^{\lambda}d\lambda\left(\frac{\bar{\theta}}{\theta}\right)^{2}\right).
\label{soln2}
\eeq

Note that the SECs 
 (\ref{sec}) and (\ref{sec2}) were analyzed in the HDC without the cosmological constant in Refs.~\cite{hong08,hong11}. 
In Section~3 just below, 
 these SECs are 
also exploited in the HDC possessing the cosmological constant. 
As soon as 
$\theta(0)$ is assumed to be negative, 
the congruence can be initially converging. The EoS inequality 
 (\ref{soln2})  
then implies that $\theta(\tau)$ should pass through the singularity within an affine length~\cite{hong08,hong11}
\beq
\lambda\leq\frac{D-2}{|\theta(0)|}+\frac{D-2}{D-1}\int_{0}^{\lambda}d\lambda\left(\frac{\bar{\theta}}{\theta}\right)^{2}.
\label{lambda2}
\eeq 
Note that the SECs 
(\ref{sec}) and 
(\ref{sec2}) are derived by using the Raychaudhuri-type equations 
 (\ref{raych}) and (\ref{raych2}) 
for the fluids of massive and massless stringy extended  objects, respectively. 
For more details of the SECs and the Raychaudhuri-type equations associated with Equations 
(\ref{sec}) and (\ref{sec2}), see Refs.~\cite{hong08,hong11}.
Note also that the SECs 
 (\ref{sec}) and 
(\ref{sec2})
~and the Raychaudhuri-type equations 
(\ref{raych}) and (\ref{raych2}) associated with the attractive gravitational forces  
are discussed in Sections 3 and 4 below.

\section{Energy Conditions and EoS Parameters for Fluid of Massive Stringy Extended Objects in Background with Cosmological Constant}\label{sec3}
\setcounter{equation}{20}

In this Section, we investigate the SECs and the ensuing EoS parameters in the HDC with the cosmological constant. 
In the results of Refs.~\cite{hong08,hong11}, we found the SECs for the fluids of both massive and massless stringy extended objects. 
Note that these SECs produced the same EoS inequalities in the HDC without the background of cosmological constant. 
In this Section, we extend these SECs to the case possessing the cosmological constant. To this end, we first 
study the stringy SECs 
(\ref{sec}) and (\ref{sec2}) by considering the energy-momentum tensor $T_{ab}$ for a perfect fluid 
in the HDC possessing the cosmological constant
:
\beq
T_{ab}=\rho u_{a}u_{b}+P(g_{ab}+u_{a}u_{b}),
\label{perfectfluid}
\eeq
where $u^{a}=(-1, 0,...,0)$ is a time-like $D$ velocity in the rest frame. Here $\rho$ and $P$ are the density and pressure of the fluid of stringy 
extended objects respectively. Combining Equations 
(\ref{eineq}) and (\ref{perfectfluid}), together with $\rho$ and $P$, one finds 
\beq
R_{ab}-\frac{1}{2}g_{ab}R=8\pi~{\rm diag}(\rho+\rho_{\Lambda}, P+P_{\Lambda}, P+P_{\Lambda}, P+P_{\Lambda}, \cdots),
\eeq
where $\rho_{\Lambda}$ and $P_{\Lambda}$ are the contributions from the cosmological constant, defined as 
\beq
\rho_{\Lambda}\equiv \frac{\Lambda}{8\pi}
\ \ \text{and}\ \ P_{\Lambda}\equiv -\frac{\Lambda}{8\pi},
\label{rholam1}
\eeq
respectively, and the ellipsis denotes the higher extended $p$-brane $(p\geq 1)$ contributions.

On the other hand, taking the trace of $T_{ab}$ 
 (\ref{perfectfluid}), one finds
\beq
T\equiv g^{ab}T_{ab}=-\rho+(D-1)P
\label{tracet}
\eeq
which, together with Equation (\ref{eineq}), yields
\beq
R\equiv g^{ab}R_{ab}=-\frac{2}{D-2}(8\pi T-D\Lambda).
\label{tracer}
\eeq
Inserting $T$ 
(\ref{tracet}) and $R$ 
 (\ref{tracer}) into Equation (\ref{eineq}), one obtains 
\beq
R_{ab}=8\pi \left(T_{ab}-\frac{1}{D-2}g_{ab}T\right)+\frac{2}{D-2}g_{ab}\Lambda.
\label{rabapp}
\eeq
For the massive stringy extended  particle, one finds 
$T_{ab}\xi^{a}\xi^{b}=\rho$ and $T_{ab}\zeta^{a}\zeta^{b}=P$ to~yield
\beq
T_{ab}(\xi^{a}\xi^{b}-\zeta^{a}\zeta^{b})=\rho-P.
\label{tabxizetaapp}
\eeq

Exploiting the SEC (\ref{sec}) for the fluid of massive stringy extended  particles 
and the Einstein field equation (\ref{eineq}) having the cosmological constant 
one finds that the SEC~produces 
\beq
R_{ab}(\xi^{a}\xi^{b}-\zeta^{a}\zeta^{b})=8\pi\left(\frac{D-4}{D-2}\rho+\frac{D}{D-2}P-\frac{1}{2\pi(D-2)}\Lambda\right)\geq 0,
\label{rab1}
\eeq
where 
Equations (\ref{tracet}), (\ref{rabapp}) and (\ref{tabxizetaapp}) are used. 
Note that in Equation (\ref{rab1}), the cosmological constant $\Lambda$ is incorporated into the results 
of Refs.~\cite{hong08,hong11}. 
Using $\rho_{\Lambda}$ and $P_{\Lambda}$ from Equation 
(\ref{rholam1}), we rewrite the EoS inequality 
(\ref{rab1})  
for the massive stringy extended  particle with cosmological constant as
\beq
R_{ab}(\xi^{a}\xi^{b}-\zeta^{a}\zeta^{b})=8\pi\left(\frac{D-4}{D-2}(\rho+\rho_{\Lambda})+\frac{D}{D-2}(P+P_{\Lambda})\right)\geq 0,
\label{rab1lamb}
\eeq
which is listed in Table 1. 
Note that in 
  Table 1, we list the EoS inequalities for the massive and massless 
 objects in 
the $D$ dimensions while in
~contrast, in Table 2, we list the EoS inequalities for the massive and massless 
point particles in the four dimensions. 
Note also 
that the massive and 
massless objects are studied in this Section and in Section 4, 
respectively.
~Note that the EoS inequality 
in Equation (\ref{rab1lamb}) yields the corresponding EoS parameter
\beq
w=\frac{P+P_{\Lambda}}{\rho+\rho_{\Lambda}}\geq -\frac{D-4}{D}.
\label{wEOSmassived}
\eeq

\begin{table}[t]
\caption{The energy conditions for fluid of stringy extended objects in the HDC defined in $D$ $(D\geq 5)$ dimensional manifold. Here we have the cosmological constant contributions $\rho_{\Lambda}=\frac{\Lambda}{8\pi}$ and $P_{\Lambda}=-\frac{\Lambda}{8\pi}$ in the strong energy conditions. Note that the EoS inequalities for fluid of massive stringy extended  objects are the same as those for fluid of massless ones as shown in the last column, even though their corresponding formulas are different to each other as shown in the second and third columns.}
\begin{center}
\begin{tabular}{llll}
\hline
Energy Condition &~~Massive, Extended Objects &~~Massless, Extended Objects &~~EoS Inequality\\
\hline
Strong &~~$R_{ab}(\xi^{a}\xi^{b}-\zeta^{a}\zeta^{b})\geq 0$ 
         &~~$R_{ab}(k^{a}k^{b}-\zeta^{a}\zeta^{b})\geq 0$ &~~$\frac{D-4}{D-2}(\rho+\rho_{\Lambda})+\frac{D}{D-2}(P+P_{\Lambda})\geq 0$\\
Weak &~~$T_{ab}(\xi^{a}\xi^{b}-\zeta^{a}\zeta^{b})\geq 0$  
         &~~$T_{ab}(k^{a}k^{b}-\zeta^{a}\zeta^{b})\geq 0$ &~~$\rho-P\geq 0$\\
\hline
\end{tabular}
\end{center}
\label{tableenergy1}
\end{table}

Note that one cannot have
 the condition $w=0$ given 
in the four dimensions, 
since the EoS parameter 
 (\ref{wEOSmassived}) has been evaluated 
in $D$ $(D\geq 5)$ dimensions in the HDC. 
Note also that the EoS parameter 
(\ref{wEOSmassived}) is not the same as that for the massive stringy extended  particle without the cosmological 
constant~\cite{hong08,hong11}, 
since Equation (\ref{wEOSmassived}) has the cosmological constant~contribution.

We investigate the relations between the EoS parameter for the fluid of stringy extended objects in the HDC defined in $D$ $(D\geq 5)$ dimensions, 
and the Hawking--Penrose limit in the massive point particle in the 
four dimensions. 
These relations were not elucidated even in the case without the cosmological constant analyzed in Refs.~\cite{hong08,hong11}. 
First, in the HDC defined in $D$ $(D\geq 5)$ dimensions, we split the SEC contributions related with 
$R_{ab}(\xi^{a}\xi^{b}-\zeta^{a}\zeta^{b})$ 
(\ref{rab1lamb})  
into three parts: (i) 
the contribution without $R_{ab}\zeta^{a}\zeta^{b}$, 
(ii)~contribution from cosmological constant $\Lambda$, and 
(iii) 
contribution associated with $R_{ab}\zeta^{a}\zeta^{b}$ only. For the first contribution (that without $R_{ab}\zeta^{a}\zeta^{b}$) which is 
related with $R_{ab}\xi^{a}\xi^{b}$ only in the background with cosmological constant, we construct
\beq
R_{ab}\xi^{a}\xi^{b}=8\pi\left(\frac{D-3}{D-2}(\rho+\rho_{\Lambda})+\frac{D-1}{D-2}(P+P_{\Lambda})\right),
\label{rabxixi}
\eeq
where $\rho$ and $P$ are the density and pressure defined in the $D$ $(D\geq 5)$ dimensions. 
Here, $\rho_{\Lambda}$ and $P_{\Lambda}$ are given in Equation (\ref{rholam1}).

Second, we obtain the contribution related with $R_{ab}\zeta^{a}\zeta^{b}$ only 
\beq
R_{ab}\zeta^{a}\zeta^{b}=8\pi\left(\frac{1}{D-2}(\rho+\rho_{\Lambda})-\frac{1}{D-2}(P+P_{\Lambda})\right).
\label{rabzetazeta}
\eeq
which, together with $R_{ab}\xi^{a}\xi^{b}$ 
(\ref{rabxixi}), reproduces Equation 
(\ref{rab1lamb}). Next, we define $R_{ab}\zeta^{a}\zeta^{b}$ 
(\ref{rabzetazeta})  
as 
\beq
R_{ab}\zeta^{a}\zeta^{b}\equiv 
-8\pi\left(\frac{D-3}{D-2}\rho_{\rm ext}+\frac{D-1}{D-2}P_{\rm ext}\right),
\label{rabzetazeta2}
\eeq
in terms of $\rho_{\rm ext}$ and $P_{\rm ext}$:  
\beq
\rho_{\rm ext}\equiv -\frac{1}{D-3}(\rho+\rho_{\Lambda})
\ \text{and}\ P_{\rm ext}\equiv \frac{1}{D-1}(P+P_{\Lambda}).
\label{rholam12}
\eeq
Combining Equations (\ref{rabxixi}) and (\ref{rabzetazeta2}), one obtains 
\beq
R_{ab}(\xi^{a}\xi^{b}-\zeta^{a}\zeta^{b})=8\pi\left(\frac{D-3}{D-2}(\rho+\rho_{\Lambda}+\rho_{\rm ext})
+\frac{D-1}{D-2}(P+P_{\Lambda}+P_{\rm ext})\right)\geq 0,
\label{rabfinal}
\eeq
which yields the effective EoS parameter
\beq
w_{0}^{\Lambda,ext}=\frac{P+P_{\Lambda}+P_{\rm ext}}{\rho+\rho_{\Lambda}+\rho_{\rm ext}}\geq -\frac{D-3}{D-1}.
\label{w1}
\eeq
Here,   
the subscript $0$ denotes $R_{ab}\zeta^{a}\zeta^{b}=0$ associated with $\rho$ and $P$. The EoS 
parameter 
(\ref{w1}) is the most general
form for the fluid of massive stringy extended objects in the background with dark energy.

Third, we study the $\rho_{\rm ext}=P_{\rm ext}=0$ case. In this case, exploiting Equation (\ref{rabzetazeta2}) 
one has 
$R_{ab}\zeta^{a}\zeta^{b}=0$ to yield $\zeta^{a}=0$. We thus find the 
four-dimensional manifold without $F$. Note that $F$ is  associated with the tangent vector field $\zeta^{a}$. 
Since the fluid of point particles defined in four dimensions 
has no extended object contribution, we have $\rho_{\rm ext}=P_{\rm ext}=0$ in Equation (\ref{w1}) 
to yield
\beq
w_{0}^{\Lambda_{0}}=\frac{P_{0}+P_{\Lambda_{0}}}{\rho_{0}+\rho_{\Lambda_{0}}}\geq -\frac{1}{3},
\label{wlam}
\eeq
for~the fluid of massive~point~particle~in~the background~with~cosmological constant. 
Here, $\rho_{\Lambda_{0}}$ and $P_{\Lambda_{0}}$ are given in terms of the four-dimensional cosmological constant $\Lambda_{0}$ 
as follows:
\beq
\rho_{\Lambda_{0}}\equiv \frac{\Lambda_{0}}{8\pi}
\ \ \text{and}\ \ P_{\Lambda_{0}}\equiv -\frac{\Lambda_{0}}{8\pi}.
\label{rholam0}
\eeq
Note that, in the four-dimensional cosmology, Equation 
(\ref{wlam}) produces the EoS inequality for the SEC for the fluid of massive point particles: 
\beq
\rho_{0}+\rho_{\Lambda_{0}}+3(P_{0}+P_{\Lambda_{0}})\geq 0,
\label{rabxixi22}
\eeq
which is listed in Table 2. In the point particle limit with $\Lambda_{0}=0$, the SEC $R_{ab}\xi^{a}\xi^{b}\geq 0$ for the fluid of 
massive point  particles produces the equivalent Hawking--Penrose EoS inequality defined in the   
four dimensions~\cite{wald84,hong08,hong11,hawking70,hawking73,carroll,hobson06}:
\beq
\rho_{0}+3P_{\Lambda_{0}}\geq 0.
\label{waldmassivein}
\eeq

\begin{table}[t]
\caption{The energy conditions for fluid of point particles possessing the mass-energy density $\rho_{0}$ and the pressure $P_{0}$ 
described in the four dimensions. Here the subscript $0$ denotes fluid of point particle, and we have the cosmological constant contributions 
$\rho_{\Lambda_{0}}=\frac{\Lambda_{0}}{8\pi}$ and $P_{\Lambda_{0}}=-\frac{\Lambda_{0}}{8\pi}$ in the strong energy conditions.}
\begin{center}
\begin{tabular}{lllll}
\hline
Energy Condition &~~Massive, Point &~~EoS Inequality  &~~Massless, Point &~~EoS Inequality\\
\hline
Strong &~~$R_{ab}\xi^{a}\xi^{b}\geq 0$ &~~$\rho_{0}+\rho_{\Lambda_{0}}+3(P_{0}+P_{\Lambda_{0}})\geq 0$
         &~~$R_{ab}k^{a}k^{b}\geq 0$ &~~$\rho_{0}+\rho_{\Lambda_{0}}+P_{0}+P_{\Lambda_{0}}\geq 0$\\
Weak &~~$T_{ab}\xi^{a}\xi^{b}\geq 0$ &~~$\rho_{0}\geq 0$
         &~~$T_{ab}k^{a}k^{b}\geq 0$ &~~$\rho_{0}+P_{0}\geq 0$\\
\hline
\end{tabular}
\end{center}
\label{tableenergy2}
\end{table}

One 
can apply the fiber bundle formalism~\cite{hong08,hong11,steenrod51} to a WEC for the fluid of massive stringy extended  objects in the 
HDC defined in $D$ ($D\geq 5$) dimensional manifold. 
 To 
 this end, let us 
assume a WEC for the fluid of massive stringy extended  objects: 
\beq
T_{ab}(\xi^{a}\xi^{b}-\zeta^{a}\zeta^{b})\geq 0.
\label{weakcarrdef}
\eeq
Note that in Equation 
(\ref{weakcarrdef}), we replace
$R_{ab}$ for the SEC with $T_{ab}$ for the WEC. Similar to the SEC case which is obtained from the 
the Raychaudhuri-type equation 
(\ref{raych})~\cite{hong08,hong11}, to construct the WEC we include 
the factor $\xi^{a}\xi^{b}-\zeta^{a}\zeta^{b}$ where $\xi^{a}$ is a time-like vector field and 
$\zeta^{a}$ is a space-like one, respectively. Exploiting $T_{ab}\xi^{a}\xi^{b}=\rho$ and $T_{ab}\zeta^{a}\zeta^{b}=P$ in 
Equation (\ref{weakcarrdef}), 
one arrives at the WEC for the massive stringy extended  particle in the HDC:
\beq
\rho-P\geq 0,
\label{weakcarr1}
\eeq
which is listed in Table 1. Note that Equation (\ref{weakcarr1}) yields the corresponding EoS parameter for 
the massive stringy extended  particle in the HDC: 
\beq
w\leq 1.
\label{weceos}
\eeq
Note also that the WEC 
(\ref{weceos}) contains the extended object contribution associated with $\zeta^{a}$.

Keeping $T_{ab}\zeta^{a}\zeta^{b}=P$, 
one obtains 
the WEC for the massive point particle in the four dimensions: 
\beq
T_{ab}\xi^{a}\xi^{b}\geq 0,
\label{weakcarr2}
\eeq
to yield the EoS inequality for the WEC for the massive point particle~\cite{wald84,hawking70,hawking73,carroll,hobson06}
\beq
\rho_{0}\geq 0,
\label{weakcarr3}
\eeq
which is listed in Table 2 and is consistent with the earlier findings
~\cite{wald84,hawking70,hawking73,carroll,hobson06}. This result suggests that the assumption 
(\ref{weakcarrdef}) is physically well defined. 
Note that the EoS inequality (\ref{weakcarr3}) for the massive point particle 
is different from  the EoS inequality (\ref{weakcarr1}) 
for the massive stringy extended particle. 

Next, in the four-dimensional cosmology without the cosmological constant, we assume 
that $\theta=\bar{\theta}=\zeta^{a}=0$ in Equation (\ref{raych}) to produce
\beq
\frac{d\theta}{d\tau}=-R_{ab}\xi^{a}\xi^{b}=-4\pi(\rho_{0}+3P_{0})
\eeq
which, using the SEC 
(\ref{waldmassivein}), 
becomes negative. The SEC for the massive point particle thus 
suggests that gravitation is attractive~\cite{carroll}. So far, 
we investigated the massive stringy extended objects and point particles. 
In the following Section~4, 
we 
 study the massless stringy extended  and point particles.

\section{Energy Conditions and EoS Parameters for Fluid of Massless Stringy Extended Objects in Background with Cosmological Constant}\label{sec4}
\setcounter{equation}{46}

In this Section, we investigate the energy conditions for the massless stringy extended objects and point particles, as well as the corresponding EoS 
parameters in the background of cosmological constant in the HDC. 
 To 
 this end, 
we first define the null vector $k^{a}$ and its auxiliary null vector $l^{a}$ as
$k^{a}\equiv \frac{1}{\sqrt{2}}(u^{a}+v^{a})$ and $l^{a}\equiv \frac{1}{\sqrt{2}}(u^{a}-v^{a})$. 
Here, $v^{a}$ is a vector residing on the subspace 
$N_{\perp}^{D-2}=\{v^{a}|v^{a}k_{a}=0,~v^{a}l_{a}=0\}$ which is a $(D-2)$ dimensional manifold 
perpendicular to the light-cone originated from $k^{a}$ and $l^{a}$. Note that the null vector field $k^{a}=(\pa/\pa \lambda)^{a}$ is given in terms of 
the affine parameter $\lambda$ and $l^{a}$ points in the opposite spatial direction to $k^{a}$. To be more specific, 
$k^{a}$ and $l^{a}$ are normalized as follows~\cite{carroll,hong11}: 
\beq
k^{a}k_{b}=l^{a}l_{a}=0 \ \ {\rm and}\ \ k^{a}l_{a}=-1.
\label{kkllka}
\eeq

For the massless stringy extended  particle, exploiting $T_{ab}$
 (\ref{perfectfluid}), we find $T_{ab}k^{a}k^{b}=\frac{1}{2}(\rho+P)$ and 
$T_{ab}\zeta^{a}\zeta^{b}=P$ to yield the EoS inequality for the WEC
\beq
T_{ab}(k^{a}k^{b}-\zeta^{a}\zeta^{b})=\frac{1}{2}(\rho-P),
\label{tabkkzetaapp}
\eeq 
which is listed in Table 1. Note that the WEC 
(\ref{tabkkzetaapp}) is for the first time 
defined for the massless stringy extended  
particle in this paper. 
Making use of the SEC (\ref{sec2})  
for the fluid of massless stringy extended  particles 
and the Einstein field equation 
(\ref{eineq}) possessing the cosmological constant, one finds 
that the SEC associated with the corresponding null vector $k^{a}$ yields
\beq 
R_{ab}(k^{a}k^{b}-\zeta^{a}\zeta^{b})=4\pi\left(\frac{D-4}{D-2}\rho+\frac{D}{D-2}P-\frac{1}{2\pi(D-2)}\Lambda\right)\geq 0.
\label{rab2}
\eeq
where 
(\ref{tracet}), (\ref{rabapp}) and (\ref{tabkkzetaapp}) are exploited. 
Note that even though we include the cosmological constant $\Lambda$ in both the massive and massless stringy extended  particle cases, the EoS 
inequality condition associated with the 
corresponding SECs 
(\ref{rab2})  
is the same as the SECs 
(\ref{rab1}). Exploiting 
$\rho_{\Lambda}$ and $P_{\Lambda}$ from Equation 
(\ref{rholam1}), we rewrite the EoS inequality 
(\ref{rab2}) for the 
massless stringy extended  particle in the cosmological constant background as
\beq
R_{ab}(k^{a}k^{b}-\zeta^{a}\zeta^{b})=4\pi\left(\frac{D-4}{D-2}(\rho+\rho_{\Lambda})+\frac{D}{D-2}(P+P_{\Lambda})\right)\geq 0,
\label{rab2lamb}
\eeq
which is listed in Table 1. Note that the EoS inequality (\ref{rab2lamb}) for the massless stringy extended  particle 
is equivalent to the inequality (\ref{rab1lamb}) 
for the massive stringy extended  particle. 
Next,  
the  EoS inequality 
(\ref{rab2lamb}) produces the EoS parameter $w$ for the massless~stringy extended ~particle~in~the  background~with~cosmological 
constant:
\beq
w=\frac{P+P_{\Lambda}}{\rho+\rho_{\Lambda}}\geq -\frac{D-4}{D}.
\label{wEOSmasslessed}
\eeq
Note that the EoS parameter $w$ 
(\ref{wEOSmasslessed}) is the same as $w$ 
(\ref{wEOSmassived}) for the massive stringy extended  particle. 
Note also that $w$ (\ref{wEOSmasslessed}) is equivalent to $w$ 
of the case for the massive stringy extended  particle without the 
cosmological constant~\cite{hong08,hong11}.

Next, in Section 3 above 
and in the current Section we treat the fluids of massless and massive stringy extended objects separately. 
However, in the real universe, these components coexist. Even if the individual EoS inequalities (bounds) are the same ($w\geq -(D-4)/D$), 
the effective equation of state of the universe ({$w_{\rm total}$) evolves as the ratio $\rho_{\rm rad}/\rho_{\rm matt}$ 
of radiation density to matter density changes. Now, we study the coexistence of the fluid of radiation (or fluid of massless 
stringy extended  objects) and the fluid of matter (or fluid of massive stringy extended  objects), and the corresponding total EoS parameter 
$w_{\rm total}$.   
 To 
 this end, 
we define $w_{\rm total}$ in terms of the ratio of radiation density to matter density
\beq
w_{\rm total}\equiv \frac{P_{\rm total}}{\rho_{\rm total}}=\frac{P_{\rm rad}+P_{\rm matt}}{\rho_{\rm rad}+\rho_{\rm matt}}.
\label{coexistence}
\eeq 
Note that our result  (\ref{coexistence})
defines the allowed theoretical region for each component in higher dimensions. Note also that the dominance of species shifts 
(which is the standard definition of a phase transition in cosmology) even if the fundamental bounds derived from
stringy geometry remain constant.  

It is appropriate to comment on the physical implications of the derived EoS inequalities related with the SECs in the HDC. 
First, as summarized in Table 1, the SECs for the fluid of stringy extended objects are defined in $D$ ($D\geq 5$) dimensional manifold. 
In other words, the EoS parameter for the fluid of massive stringy extended objects is obtainable from 
$R_{ab}(\xi^{a}\xi^{b}-\zeta^{a}\zeta^{b})$ 
(\ref{rab1lamb}). 
Similarly, one can obtain the EoS parameter for the fluid of massless stringy extended objects from $R_{ab}(k^{a}k^{b}-\zeta^{a}\zeta^{b})$ 
(\ref{rab2lamb}). 
Note that the expressions $R_{ab}(\xi^{a}\xi^{b}-\zeta^{a}\zeta^{b})$ and $R_{ab}(k^{a}k^{b}-\zeta^{a}\zeta^{b})$ 
possess the non-vanishing tangent vector field $\zeta^{a}$ which resides on the fiber space for the stringy extended  object having finite size.

Second, in $D=5$ we find the EoS parameter $w\geq -1/5$, which 
becomes negative allowing for negative pressure. In higher dimensions, this feature of the negative pressure means for exotic matter, the aspect of 
which is similar to the standard point particle limit related with the region $-1/3\leq w<0$. In general, for the fluid of stringy extended  objects 
defined in $D$ ($D\geq 5$) dimensions, the EoS parameter bound becomes negative, allowing for negative pressure. Note that, as summarized in Table 2, 
the SECs for the fluid of point particles are defined in the four-dimensional manifold without $F$ associated with $\zeta^{a}$ field.

We study the relations between the EoS parameter $w$ 
(\ref{wEOSmasslessed}) and the Hawking--Penrose limit in the fluid of massless point 
particles in the four dimensions}. 
Similar to the fluid of massive stringy extended object case discussed in Section 3, 
we split the SEC contributions associated with $R_{ab}(k^{a}k^{b}-\zeta^{a}\zeta^{b})$ 
(\ref{rab2lamb})  
into three pieces: (i) 
contribution without $R_{ab}\zeta^{a}\zeta^{b}$, (ii) 
contribution from cosmological constant $\Lambda$, and 
 (iii) 
contribution related with $R_{ab}\zeta^{a}\zeta^{b}$ only. First, similar to Equation 
(\ref{rabxixi}), for the first contribution 
(that without $R_{ab}\zeta^{a}\zeta^{b}$) which is associated  with $R_{ab}k^{a}k^{b}$ only in the background with cosmological constant, 
one obtains
\beq
R_{ab}k^{a}k^{b}=4\pi\left(\rho+\rho_{\Lambda}+P+P_{\Lambda}\right),
\label{rabkk}
\eeq
where $\rho$ and $P$ are the density and pressure defined in the $D$ $(D\geq 5)$ dimensions. 
Here,  $\rho_{\Lambda}$ and $P_{\Lambda}$ are defined in Equation (\ref{rholam1}).

Second, we find the contribution of the fluid of massless objects associated with the vector field $\zeta^{a}$ as follows
\beq
R_{ab}\zeta^{a}\zeta^{b}=8\pi\left(\frac{1}{D-2}(\rho+\rho_{\Lambda})-\frac{1}{D-2}(P+P_{\Lambda})\right).
\label{rabzetazeta22}
\eeq
Note that, together with $R_{ab}k^{a}k^{b}$ 
(\ref{rabkk}), Equation (\ref{rabzetazeta22}) reproduces Equation (\ref{rab2lamb}). 

Third, we define $R_{ab}\zeta^{a}\zeta^{b}$ in Equation (\ref{rabzetazeta22}) as
\beq
R_{ab}\zeta^{a}\zeta^{b}\equiv 
-4\pi\left(\rho_{\rm ext}+P_{\rm ext}\right),
\label{rabzetazeta3}
\eeq
in terms of $\rho_{\rm ext}$ and $P_{\rm ext}$:  
\beq
\rho_{\rm ext}\equiv -\frac{2}{D-2}(\rho+\rho_{\Lambda})\ \ {\rm and} \ \ P_{\rm ext}\equiv \frac{2}{D-2}(P+P_{\Lambda}).
\label{rholam122}
\eeq
Note that 
$\rho_{\rm ext}$ and $P_{\rm ext}$ in Equation (\ref{rholam122}) 
for the fluid of massless stringy extended  objects are different from 
those for the fluid of massive ones defined in Equation (\ref{rholam12}).

Combining (\ref{rabkk}) and (\ref{rabzetazeta3}), one finds
\beq
R_{ab}(k^{a}k^{b}-\zeta^{a}\zeta^{b})=4\pi\left(\rho+\rho_{\Lambda}+\rho_{\rm ext}+P+P_{\Lambda}+P_{\rm ext}\right)\geq 0,
\label{rabfinal2}
\eeq
which yields the effective EoS parameter  
\beq
w_{0}^{\Lambda, ext}=\frac{P+P_{\Lambda}+P_{\rm ext}}{\rho+\rho_{\Lambda}+\rho_{\rm ext}}\geq -1.
\label{w2}
\eeq
Here, the subscript $0$ denotes $R_{ab}\zeta^{a}\zeta^{b}=0$ associated with $\rho$ and $P$. The EoS parameter 
(\ref{w2}) is the most general form for 
the fluid of massless stringy extended  objects~in~the background with~dark~energy.

We investigate the fluid of four-dimensional point particle limit of the fluid of massless stringy extended objects. 
Since the fluid of point particles have no extended object contribution, we have $\rho_{\rm ext}=P_{\rm ext}=0$ in Equation (\ref{w2}) to yield
\beq
w_{0}^{\Lambda_{0}}=\frac{P_{0}+P_{\Lambda_{0}}}{\rho_{0}+\rho_{\Lambda_{0}}}\geq -1,
\label{wlambda22}
\eeq
for the~fluid of massless~point~particles~in~the background~with~cosmological constant. 
Here, $\rho_{\Lambda_{0}}$ and $P_{\Lambda_{0}}$ are given in Equation  
(\ref{rholam0}). Note that Equation (\ref{wlambda22}) yields the EoS inequality for the SEC for the fluid of massless point particles in 
the background~with~cosmological constant~\cite{wald84,hawking70,hawking73,carroll,hobson06}:
\beq
\rho_{0}+\rho_{\Lambda_{0}}+P_{0}+P_{\Lambda_{0}}\geq 0,
\label{wlambda222}
\eeq
which is listed in Table 2. Moreover, in the point particle limit with $\Lambda_{0}=0$, 
the SEC $R_{ab}k^{a}k^{b}\geq 0$ for the fluid of massless point particles yields the EoS inequality defined in the four-dimensional 
cosmology~\cite{wald84,hawking70,hawking73,carroll,hobson06}:
\beq
\rho_{0}+P_{\Lambda_{0}}\geq 0.
\label{waldmasslessin}
\eeq


One 
can apply the fiber bundle formalism~\cite{hong08,hong11,steenrod51} to a WEC for the fluid of massless stringy extended  objects 
in the HDC defined in $D$ ($D\geq 5$) dimensional manifold. 
 To 
 this end, 
we assume a WEC for the fluid of massive stringy extended  objects: 
\beq
T_{ab}(k^{a}k^{b}-\zeta^{a}\zeta^{b})\geq 0.
\label{weakcarrdef2}
\eeq
Note that in Equation 
(\ref{weakcarrdef2}) 
$R_{ab}$ for the SEC is replaced with $T_{ab}$ for the WEC. Similar to the SEC case which is obtained from the 
the Raychaudhuri-type equation 
(\ref{raych2})~\cite{hong08,hong11}, to construct the WEC we included 
the factor $\xi^{a}\xi^{b}-\zeta^{a}\zeta^{b}$ where $\xi^{a}$ is a time-like vector field and 
$\zeta^{a}$ is a space-like one, respectively. Using $T_{ab}k^{a}k^{b}=\frac{1}{2}(\rho+P)$ and $T_{ab}\zeta^{a}\zeta^{b}=P$ in 
Equation (\ref{weakcarrdef2}), 
one ends 
 up with the WEC for the fluid of massless stringy extended  objects in the HDC: 
\beq
\rho-P\geq 0,
\label{weakcarr12}
\eeq
which is listed in Table 1.

Next we define the WEC for the fluid of massless point particles 
(or null energy condition) in the four-dimensional cosmology: 
\beq
T_{ab}k^{a}k^{b}\geq 0.
\label{weakcarr22}
\eeq
The WEC 
 (\ref{weakcarr22})  
produces~\cite{wald84,hawking70,hawking73,carroll,hobson06}
\beq
\rho_{0}+P_{0}\geq 0,
\label{weakcarr32}
\eeq
which is listed in Table 2. Note that, in the HDC, the WEC (\ref{weakcarr12})
for the fluid of massless stringy extended  objects 
is  the same as the WEC (\ref{weakcarr1}) 
for the massive stringy extended  objects. 
Note that, in the WECs, there is no 
effect of the background with cosmological constant since the 
definitions in Equations (\ref{weakcarrdef}) and (\ref{weakcarrdef2}) do not possess the corresponding cosmological constant terms. 
Similar to Equation (\ref{weakcarr3}) for the WEC inequality for the fluid of massive point particles, in the four-dimensional cosmology, we find 
the WEC inequality (\ref{weakcarr32}) 
for the fluid of massless point particles 
which are consistent with the results obtained in 
Refs.~\cite{wald84,hawking70,hawking73,carroll,hobson06}. This finding 
also suggests that the assumption made in  Equation (\ref{weakcarrdef2}) is physically well 
defined.
Note in Equations 
(\ref{wEOSmassived}) and (\ref{wEOSmasslessed}) that, in the radiation-dominated and matter-dominated eras in the background with cosmological 
constant, one finds 
the same EoS parameter $w$ in the SEC in the HDC. 
The stringy SECs thus impose a universal constraint on the EoS parameter $w$ that remains valid across both radiation- and matter-dominated eras. 
This feature is the same as the case without the cosmological constant~\cite{hong08,hong11}.

One of the simplest models to explain the dark energy is the $\Lambda$CDM 
 model, where the dark energy is considered static and positive. 
This constant accelerates the expansion of the universe. To be more specific, the cosmological constant $\Lambda$ in Equation (\ref{sss2}) is 
non-zero and positive. 
This feature implies that the expansion of the universe is accelerating, which is consistent with the corresponding discovery in 1998~\cite{per98}. 
Note that the dark energy does not need to be constant. For instance, the quintessence model describes dynamical dark energy~\cite{li04,roy25,li25} 
having time dependent scalar field. Moreover, in the $f(R)$ theories of the gravity, the integrand in the action $S_{\rm gr}$ in 
Equation (\ref{sss2}) is 
replaced as follows: $\sqrt{-g}(R-2\Lambda)\rightarrow \sqrt{-g}f(R)$ with $f(R)$ being a function of $R$. 
We study explicitly the case that the cosmological constant is static in the HDC. In this case the energy density is given as 
$\rho_{\Lambda}=\frac{\Lambda}{8\pi}$, for~both~the fluid of massive~and~massless~stringy extended ~objects as shown in 
Equation (\ref{rholam1}). In order to explain the accelerating expansion of the universe, one needs 
to find an energy having negative pressure $P_{\Lambda}=-\frac{\Lambda}{8\pi}$ with the positive cosmological constant $\Lambda$. 
This 
 signifies 
 that the accelerating universe in the HDC possesses a positive cosmological constant.

Finally, in the four-dimensional cosmology without the background~with~cosmological constant, we assume that 
$\theta=\bar{\theta}=\zeta^{a}=0$ in (\ref{raych2}) to produce
\beq
\frac{d\theta}{d\lambda}=-R_{ab}k^{a}k^{b}=-4\pi(\rho_{0}+P_{0})
\eeq
which, using the SEC 
(\ref{weakcarr32}), becomes negative. As a result, the SEC for the fluid of massless point particles 
implies that gravitation is attractive~\cite{carroll}. 
Note that exploiting $\frac{d\theta}{d\tau}=-R_{ab}(\xi^{a}\xi^{b}-\zeta^{a}\zeta^{b})\leq 0$ and
$\frac{d\theta}{d\lambda}=-R_{ab}(k^{a}k^{b}-\zeta^{a}\zeta^{b})\leq 0$ in the HDC having $\theta=\bar{\theta}=0$ and $\zeta^{a}\neq 0$, 
the SECs for the fluids of massive and massless stringy extended  objects suggest that gravitation is attractive.

\section{Conclusions}\label{sec5}

In summary, we have 
investigated the energy conditions in the HDC having a cosmological constant $\Lambda$. As shown in Table 1, 
the EoS inequalities for the SECs and WECs for the fluid of massive stringy extended  objects are equivalent to those for fluid of massless 
stringy extended  objects in the HDC. In contrast, in the four-dimensional cosmology considered 
in Section 3, we evaluated the EoS inequalities for the SECs and  
WECs for fluids of massive and massless point particles as listed in Table 2. 
Next, we studied the EoS parameter $w$ in the HDC 
with the cosmological constant. To be specific, we constructed $w\geq -(D-4)/D$ for both the fluids of massive and massless stringy extended  objects 
in the HDC 
in the background~with~cosmological constant. We thus concluded that the stringy SECs impose a universal constraint 
on $w$ that remains valid across both radiation- and matter-dominated eras.
Moreover, we split the EoS parameter contributions for the stringy extended  objects in the background with cosmological constant as $w_{0}^{\Lambda, 
ext}=(P+P_{\Lambda}+P_{\rm ext})/
(\rho+\rho_{\Lambda}+\rho_{\rm ext})$ where $(\rho, P)$, $(\rho_{\Lambda}, P_{\Lambda})$ and 
$(\rho_{\rm ext}, P_{\rm ext})$ are contribution from the contribution without $R_{ab}\zeta^{a}\zeta^{b}$, contribution from cosmological constant 
$\Lambda$, and contribution 
with $R_{ab}\zeta^{a}\zeta^{b}$ only, respectively. 
Here, $\zeta^{a}$ is the tangent vector field associated with the fiber space in the HDC. 
Since the point particles have no extended object contribution, then 
$\rho_{\rm ext}=P_{\rm \rm ext}=0$. In the point particle limit with 
$\Lambda=0$, 
the SEC produces the EoS parameter related with the Hawking--Penrose-type EoS inequality. 
Moreover, the effect of the background with cosmological constant was shown to not exist in the WECs. 
This is one of the main points of this paper. 
It is of interest 
 to investigate the $f(R)$ gravity theory having the Gauss--Bonnet term 
$R_{GB}^{2}=R^{2}-4R_{ab}R^{ab}+R_{abcd}R^{abcd}$~\cite{bhl23,bhl24} in the HDC. The next intiguing 
topic is the FLRW model in the HDC in which we investigate the higher dimensional metric $g_{ab}$ $(a,b=0,1,2,3,4,...)$ of the form
~$g_{ab}={\rm diag}(-1,a^{2},a^{2},a^{2},b^{2},\cdots)$,
where now 
$a$ is a scale factor defined in the four-dimensional base manifold and $b$ is a scale factor residing on the compact extended $(p=1)$-brane, 
and the ellipsis stands for the higher extended $p$-brane $(p\geq 2)$ contributions, respectively. On the other hand, it 
would be of interest to investigate 
the wormhole geometry~\cite{hobson06,para20,bhl242} and the corresponding energy conditions and EoS parameters in the HDC. 
The next 
topic to address would 
 be to study the black hole and the corresponding energy conditions~\cite{ion25,bah24,bal14} 
 and EoS parameters in the HDC.

\acknowledgments{B.-H.L.
~was supported by the National Research Foundation of Korea (NRF)  (RS-2020-NR049598, RS-2024-00339204, and RS-2026-25473640) and by Overseas Visiting 
Fellow Program of Shanghai University. S.-T.H.
~was supported by Basic Science Research Program through the NRF funded by the Ministry of Education (NRF-2019R1I1A1A01058449).
B.-H.
L. 
would like to thank the hospitality of the KIAS (Korean Institute for Advanced Study) 
 and APCTP (Asia Pacific Center for Theoretical Physics) where a part of this work has been done during the visit.}



\appendix
\section{Four-Dimensional Point Particle Cosmology with Cosmological~Constant}\label{app1}
\setcounter{equation}{0}
\renewcommand{\theequation}{A\arabic{equation}}

In this Appendix, 
we study an action for a four-dimensional gravity with the fluid of point particles which are applied to the HDC in 
Section 2.  The action for the four-dimensional gravity with point particles in the background with cosmological constant is described 
as 
\bea
S&=&S_{p=0}+S_{\rm gr}+S_{\rm pf},\nn\\
S_{p=0}&=&-m\int_{\tau_{1}}^{\tau_{2}}d\tau f(\tau),\nn\\
S_{\rm gr}&=&\frac{1}{16\pi}\int d^{3}x\sqrt{-g}(R-2\Lambda),
\label{sss}
\eea
where $S_{p=0}$, $S_{\rm gr}$ and $S_{\rm pf}$ are the relativistic action for the fluid of point particle and actions of gravity and perfect fluid 
associated with the energy-momentum tensor of the form $T_{ab}$ in (\ref{perfectfluid}). 
Here, $m$ is a test point particle mass and $f(\tau)$ is defined as~\cite{scherk75}
~$f(\tau)=\left(-g_{ab}\frac{\pa x^{a}}{\pa\tau}\frac{\pa x^{b}}{\pa\tau}\right)^{1/2}=(-g_{ab}\xi^{a}\xi^{b})^{1/2}$, where 
the notation $\xi^{a}=(\pa/\pa\tau)^{a}$ ($a=0,1,2,3$) with metric $(-,+,+,+)$ is used, and
~$\Lambda$ is the cosmological constant. 
In 
 this paper, 
 we restrict ourselves to the case that the variation of $f(\tau)$ under $\delta g^{ab}$ is neglected, for simplicity. 
This restriction is also 
applied to the stringy extended  objects in Section 2.
 In order to
define the action on the curved manifold, let $(M, g_{ab})$ be a four-dimensional manifold associated with the metric $g_{ab}$.
Given $g_{ab}$, there is 
a unique covariant derivative
$\na_{a}$ satisfying~\cite{wald84} $\na_{a}g_{bc}=0$,
$\na_{a}\omega^{b}=\pa_{a}\omega^{b}+\Gamma^{b}_{~ac}~\omega^{c}$
and $(\na_{a}\na_{b}-\na_{b}\na_{a})\omega_{c}=R_{abc}^{~~~d}~\omega_{d}$. 

{We introduce the deviation vector field $\eta^{a}=(\pa/\pa
\alpha)^{a}$ which represents the displacement to an
infinitesimally nearby world line, and let $\Sigma$ denote the 
two-dimensional submanifold spanned by the world lines
$\gamma_{\alpha}(\tau)$~\cite{wald84}. One then may choose $\tau$ and $\alpha$ as coordinates of $\Sigma$ to yield the
commutator relation, 
\beq
\pounds_{\xi}\eta^{a}=\xi^{b}\na_{b}\eta^{a}-\eta^{b}\na_{b}\xi^{a}=0.
\label{commpoint}
\eeq}

We construct the first derivative of $S_{p}$ with respect to $\alpha$ to produce
\beq
\frac{dS_{p=0}}{d\alpha}=-m\int_{\tau_{1}}^{\tau_{2}}d\tau \eta^{a}\na_{a}f=-m\int_{\tau_{1}}^{\tau_{2}}d\tau \eta^{a}\frac{1}{f}(-\xi^{b}\na_{a}\xi_{b})
=-\int_{\tau_{1}}^{\tau_{2}}d\tau  P_{\tau}^{b}\eta^{a}\na_{a}\xi_{b},
\label{eom1}
\eeq
where we used the energy-momentum current $P_{\tau}^{a}=-\frac{m}{f}\xi^{a}$. Exploiting the commutator 
(\ref{commpoint}), one arrives 
at
\beq
\frac{dS_{p=0}}{d\alpha}=-\int_{\tau_{1}}^{\tau_{2}}d\tau P_{\tau}^{b}\xi^{a}\na_{a}\eta_{b}
=\int_{\tau_{1}}^{\tau_{2}}d\tau \eta_{b}\xi^{a}\na_{a}P_{\tau}^{b}-P_{\tau}^{a}\eta_{a}|_{\tau=\tau_{1}}^{\tau=\tau_{2}},\label{eom10}
\eeq
and the boundary term vanishes if 
the end point condition 
\beq
\eta^{a}(\tau=\tau_{1})=\eta^{a}(\tau=\tau_{2})=0,
\label{endpoint}
\eeq
is used to yield a geodesic equation
\beq
\xi^{a}\na_{a}P_{\tau}^{b}=0.
\label{geo1a}
\eeq
Applying the time-like condition $\xi\cdot\xi=-1$ to Equation (\ref{geo1a}), 
one obtains the geodesic equation of the form
\beq
\xi^{a}\na_{a}\xi^{b}=0.
\eeq
Using the first equation in Equation (\ref{eom10}) one finds 
 the second derivative of $S_{p}$ as follows: 
\beq
\frac{d^{2}S_{p=0}}{d\alpha^{2}}=-\int_{\tau_{1}}^{\tau_{2}} d\tau\eta^{c}\na_{c}(P_{\tau}^{b}\xi^{a}\na_{a}\eta_{b})
=-\int_{\tau_{1}}^{\tau_{2}} d\tau\eta^{c}[(\na_{c}P_{\tau}^{b})\xi^{a}\na_{a}\eta_{b}+P_{\tau}^{b}\{ (\na_{c}\xi^{a})\na_{a}\eta_{b}
+\xi^{a}\na_{c}\na_{a}\eta_{b}\}].
\label{gdev10}
\eeq
Making use of the relation $\na_{c}\na_{a}\eta_{b}=\na_{a}\na_{c}\eta_{b}+R_{cab}^{~~~d}\eta_{d}$ and the commutator 
(\ref{commpoint}), one finds
\bea
\frac{d^{2}S_{p=0}}{d\alpha^{2}}&=&-\int_{\tau_{1}}^{\tau_{2}} d\tau [(\eta^{c}\na_{c}P_{\tau}^{b})\xi^{a}\na_{a}\eta_{b}
+P_{\tau}^{b}\{ 
(\xi^{c}\na_{c}\eta^{a})\na_{a}\eta_{b}+\xi^{a}\eta^{c}\na_{a}\na_{c}\eta_{b}
+R_{cab}^{~~~d}\eta^{c}\eta_{d}\xi^{a}
\}]\nn\\
&=&-\int_{\tau_{1}}^{\tau_{2}} d\tau [(\eta^{c}\na_{c}P_{\tau}^{b})\xi^{a}\na_{a}\eta_{b}
+P_{\tau}^{b}\{ 
(\xi^{c}\na_{c}\eta^{a})\na_{a}\eta_{b}+\xi^{c}\eta^{a}\na_{c}\na_{a}\eta_{b}\}
-R_{acb}^{~~~d}\eta^{c}\eta_{d}\xi^{a}P_{\tau}^{b}
]\nn\\
&=&-\int_{\tau_{1}}^{\tau_{2}} d\tau [(\eta^{c}\na_{c}P_{\tau}^{b})\xi^{a}\na_{a}\eta_{b}
+P_{\tau}^{b}(\xi^{c}\na_{c})(\eta^{a}\na_{a}\eta_{b})-R_{bcd}^{~~~a}\eta^{c}\eta_{a}\xi^{b}P_{\tau}^{d}]\nn\\
&=&-\int_{\tau_{1}}^{\tau_{2}} d\tau [\xi^{a}\na_{a}(\eta_{b}\eta^{c}\na_{c}P_{\tau}^{b})-\eta_{b}\xi^{a}\na_{a}(\eta^{c}\na_{c}P_{\tau}^{b})
+\xi^{c}\na_{c}(P_{\tau}^{b}\eta^{a}\na_{a}\eta_{b})
-(\eta^{a}\na_{a}\eta_{b})\xi^{c}\na_{c}P_{\tau}^{b}-R_{bcd}^{~~~a}\eta^{c}\eta_{a}\xi^{b}P_{\tau}^{d}]\nn\\
&=&-\eta_{b}\eta^{c}\na_{c}P_{\tau}^{b}|_{\tau=\tau_{1}}^{\tau=\tau_{2}}-P_{\tau}^{b}\eta^{a}\na_{a}\eta_{b}|_{\tau=\tau_{1}}^{\tau=\tau_{2}}
+\int_{\tau_{1}}^{\tau_{2}} d\tau [\eta_{b}\xi^{a}\na_{a}(\eta^{c}\na_{c}P_{\tau}^{b})+R_{bcd}^{~~~a}\eta^{c}\eta_{a}\xi^{b}P_{\tau}^{d}].
\label{gdev102}
\eea
Inserting the end point condition 
(\ref{endpoint}) and the geodesic equation 
(\ref{geo1a}) into Equation (\ref{gdev102}), one ends 
up with a 
geodesic deviation equation of the form
\beq
\frac{d^{2}S_{p=0}}{d\alpha^{2}}=\int_{\tau_{1}}^{\tau_{2}} d\tau\eta_{a}(\chi\eta)^{a},
\label{gdev1}
\eeq
where 
\beq
(\chi\eta)^{a}=\xi^{b}\na_{b}(\eta^{c}\na_{c}P_{\tau}^{a})+R_{bcd}^{~~~a}\xi^{b}\eta^{c}P_{\tau}^{d}.
\label{aleph1}
\eeq

On the other hand, the variation in $S_{\rm gr}+S_{\rm pf}$ under the change $\delta g^{ab}$ ($a,b=0,1,2,3$) produces the Einstein field equation
\beq
R_{ab}-\frac{1}{2}g_{ab}R+g_{ab}\Lambda=8\pi T_{ab},
\label{einfieldeq2}
\eeq
where we exploited the definition $R_{ab}=R_{acb}^{~~~c}$ with $R_{bcd}^{~~~a}$ being defined in (\ref{aleph1}). 
Here, we used the relation $\delta S_{\rm pf}/\delta g^{ab}=-\frac{1}{2}\sqrt{-g}T_{ab}$.

\section[\appendixname~\thesection]{Raychaudhuri-Type Equation in HDC}\label{app2}
\setcounter{equation}{12}

In this Appendix, 
 we construct the details of the Raychaudhuri-type equation 
(\ref{rayeqn}) more completely and rigorously in the HDC, 
since in Refs.~\cite{hong08,hong11} we stated the sketch of the Raychaudhuri-type equation. 
 To 
 this end, in the orthonormal gauge, we introduce tensor fields $B_{ab}$ and $\bar{B}_{ab}$ defined as 
\beq
B_{ab}=\na_{b}\xi_{a} \ \ {\rm and} \ \ \bar{B}_{ab}=\na_{b}\zeta_{a},
\label{baba}
\eeq
which fulfill the identities
\beq
B_{ab}\xi^{a}=0,~\bar{B}_{ab}\zeta^{a}=0  \ \ {\rm and} \ \ B_{ab}\xi^{b}-\bar{B}_{ab}\zeta^{b}=0.
\label{idena}
\eeq
Exploiting the commutator relations 
(\ref{poundxizeta}), one finds 
\beq
\xi^{a}\na_{a}\eta^{b}-\zeta^{a}\na_{a}\eta^{b}=(B^{b}_{~a}-\bar{B}^{b}_{~a})\eta^{a}.\eeq
We define the metrics $h_{ab}$ and $\bar{h}_{ab}$ as
\beq
h_{ab}=g_{ab}+\xi_{a}\xi_{b}
\ \text{and}\ \bar{h}_{ab}=g_{ab}-\zeta_{a}\zeta_{b},
\label{projections}
\eeq
to produce the identities 
\beq
h^{ab}h_{ab}=\bar{h}^{ab}\bar{h}_{ab}=D-1.
\label{iddb}
\eeq 

Using $h_{ab}$ and $\bar{h}_{ab}$ defined in Equation  
(\ref{projections}), we decompose $B_{ab}$ into three parts: 
\beq
B_{ab}=\frac{1}{D-1}\theta h_{ab}+\sigma_{ab}+\omega_{ab},
\label{bab}
\eeq 
with 
\beq
\theta=B^{ab}h_{ab},~
\sigma_{ab}=B_{(ab)}-\frac{1}{D-1}\theta h_{ab},
\ \text{and}\ \omega_{ab}=B_{[ab]},
\label{thetas}
\eeq 
and $B_{ab}$ into three pieces:  
\beq
\bar{B}_{ab}=\frac{1}{D-1}\bar{\theta}\bar{h}_{ab}+\bar{\sigma}_{ab}+\bar{\omega}_{ab},
\label{barbab}
\eeq
with 
\beq
\bar{\theta}=\bar{B}^{ab}\bar{h}_{ab},~
\bar{\sigma}_{ab}=\bar{B}_{(ab)}-\frac{1}{D-1}\bar{\theta}\bar{h}_{ab},
\ \text{and}\ \bar{\omega}_{ab}=\bar{B}_{[ab]}.
\label{barthetaab}
\eeq
One finds 
the identities
\bea
\sigma_{ab}h^{ab}&=&\omega_{ab}h^{ab}=0 \ \ \ {\rm and} \ \ \sigma_{ab}\xi^{b}=\omega_{ab}\xi^{b}=\frac{1}{2}B_{ab}\xi^{b},\nn\\
\bar{\sigma}_{ab}\bar{h}^{ab}&=&\bar{\omega}_{ab}\bar{h}^{ab}=0  \ \ {\rm and} \ \ \bar{\sigma}_{ab}\zeta^{b}=\bar{\omega}_{ab}\zeta^{b}=\frac{1}{2}\bar{B}_{ab}\zeta^{b}.
\label{idsss}
\eea
to yield $-\sigma_{ab}\xi^{b}+\bar{\sigma}_{ab}\zeta^{b}=0$, where 
the stringy geodesic equation 
(\ref{gEOSe}) is used.

We evaluate the following quantity\vspace{-12pt}
\bea
\xi^{c}\na_{c}B_{ab}&=&\xi^{c}\na_{c}\na_{b}\xi_{a}=\xi^{c}(\na_{b}\na_{c}\xi_{a}+R_{cba}^{~~~d}\xi_{d})
=\na_{b}(\xi^{c}\na_{c}\xi_{a})-(\na_{b}\xi^{c})(\na_{c}\xi_{a})+R_{cba}^{~~~d}\xi^{c}\xi_{d}\nn\\
&=&\na_{b}(\xi^{c}\na_{c}\xi_{a})-B^{c}_{~b}B_{ac}+R_{cba}^{~~~d}\xi^{c}\xi_{d}.
\label{quantitiesb}
\eea
Similarly one finds
\beq
\zeta^{c}\na_{c}\bar{B}_{ab}=\na_{b}(\zeta^{c}\na_{c}\zeta_{a})-\bar{B}^{c}_{~b}\bar{B}_{ac}+R_{cba}^{~~~d}\zeta^{c}\zeta_{d}.
\label{quantitiesb2}
\eeq
Combining the identities 
(\ref{quantitiesb}) and (\ref{quantitiesb2}), one arrives 
 at
\beq
-\xi^{c}\na_{c}B_{ab}+\zeta^{c}\na_{c}\bar{B}_{ab}=B^{c}_{~b}B_{ac}-\bar{B}^{c}_{~b}\bar{B}_{ac}-R_{cbad}(\xi^{c}\xi^{d}-\zeta^{c}\zeta^{d}),
\label{rayob}
\eeq
where 
again 
the stringy geodesic equation 
(\ref{gEOSe}) is exploited. We calculate the trace of the above equation by multiplying both sides of 
Equation (\ref{rayob}) by $g^{ab}$. Firstly, we rewrite the trace of left-hand side of Equation (\ref{rayob}) as
\beq
-\xi^{c}\na_{c}(B_{ab}g^{ab})+\zeta^{c}\na_{c}(\bar{B}_{ab}g^{ab})
=-\xi^{c}\na_{c}(B_{ab}(h^{ab}-\xi^{a}\xi^{b}))+\zeta^{c}\na_{c}(\bar{B}_{ab}(\bar{h}^{ab}+\zeta^{a}\zeta^{b}))
=-\xi^{c}\na_{c}\theta+\zeta^{c}\na_{c}\bar{\theta},
\label{ray1b}
\eeq
where Equations 
(\ref{idena}), (\ref{projections}), (\ref{thetas}) and (\ref{barthetaab}) are used. 
Secondly, we rewrite the trace of right-hand side of Equation (\ref{rayob}) as\vspace{-12pt}
\bea
&&B^{ca}B_{ac}-\bar{B}^{ca}\bar{B}_{ac}+R_{cd}(\xi^{c}\xi^{d}-\zeta^{c}\zeta^{d})
=\left(\frac{1}{D-1}\theta h^{ca}+\sigma^{ca}+\omega^{ca}\right)\left(\frac{1}{D-1}\theta h_{ac}+\sigma_{ac}+\omega_{ac}\right)\nn\\
&&-\left(\frac{1}{D-1}\bar{\theta}\bar{h}^{ca}+\bar{\sigma}^{ca}+\bar{\omega}^{ca}\right)
\left(\frac{1}{D-1}\bar{\theta}\bar{h}_{ac}+\bar{\sigma}_{ac}+\bar{\omega}_{ac}\right)
+R_{cd}(\xi^{c}\xi^{d}-\zeta^{c}\zeta^{d})\nn\\
&&=\frac{1}{D-1}(\theta^{2}-\bar{\theta}^{2})+\sigma_{ab}\sigma^{ab}-\bar{\sigma}_{ab}\bar{\sigma}^{ab}
-\omega_{ab}\omega^{ab}+\bar{\omega}_{ab}\bar{\omega}^{ab}+R_{ab}(\xi^{a}\xi^{b}-\zeta^{a}\zeta^{b}),
\label{ray2b}
\eea
where Equations 
(\ref{iddb}), (\ref{bab}), (\ref{barbab}) and (\ref{idsss}) are exploited.

Combining the identities 
(\ref{ray1b}) and (\ref{ray2b}), one is 
 left with 
\beq
-\xi^{c}\na_{c}\theta+\zeta^{c}\na_{c}\bar{\theta}=\frac{1}{D-1}(\theta^{2}-\bar{\theta}^{2})+\sigma_{ab}\sigma^{ab}-\bar{\sigma}_{ab}\bar{\sigma}^{ab}
-\omega_{ab}\omega^{ab}+\bar{\omega}_{ab}\bar{\omega}^{ab}+R_{ab}(\xi^{a}\xi^{b}-\zeta^{a}\zeta^{b}).
\label{rayfinalb}
\eeq
Exploiting 
$-\xi^{c}\na_{c}\theta+\zeta^{c}\na_{c}\bar{\theta}=-
{d\theta}/{d\tau}
+
{d\bar{\theta}}/{d\sigma}$, 
one finds that Equation (\ref{rayfinalb}) 
reproduces Equation (\ref{rayeqn}).



\end{document}